\documentclass[journal, a4paper]{IEEEtran}

\usepackage{cite}

\ifCLASSINFOpdf
  \usepackage[pdftex]{graphicx}
  \graphicspath{{fig/}}
  \DeclareGraphicsExtensions{.pdf,.jpeg,.png}
\else
\fi

\usepackage{amsmath,amsfonts,amssymb}
\usepackage{algpseudocode}
\usepackage{algorithm}
\usepackage{array}
\usepackage{multirow}
\usepackage{trfsigns}
\usepackage[ansinew]{inputenc}
\usepackage{bm}
\usepackage{subcaption}	
\usepackage{import}
\usepackage{xcolor}
\usepackage[squaren]{SIunits}
\usepackage{upgreek}
\usepackage[hidelinks]{hyperref}
\usepackage{booktabs}
\usepackage{soul}
\usepackage{wasysym}
\usepackage{cancel}
\usepackage{struktex}
\MathNormal
\usepackage{bm}
\usepackage{pifont}
\usepackage{mathtools}
\usepackage{tablefootnote}

\renewcommand{\figurename}{Fig.}
\renewcommand{\tablename}{Tab.}

\newcommand{\figref}[1]{\figurename~\ref{#1}}
\newcommand{\tabref}[1]{\tablename~\ref{#1}}

\newcolumntype{C}[1]{>{\centering\arraybackslash}p{#1}}

\begin{document}
\renewcommand\arraystretch{1.4}
\title{Technical Report on Reinforcement Learning Control on the Lucas-N\"ulle Inverted Pendulum}
\author{\IEEEauthorblockN{Maximilian Schenke, Shalbus Bukarov}
\thanks{M. Schenke is a freelance engineer based in Paderborn, Germany, S. Bukarov is with the Lucas-N\"ulle GmbH in Kerpen, Germany.
\\
\mbox{E-mail: }schenke@lea.uni-paderborn.de, shalbus.bukarov@lucas-nuelle.de}
}

\maketitle
\begin{abstract}
The discipline of automatic control is making increased use of concepts that originate from the domain of machine learning. Herein, reinforcement learning (RL) takes an elevated role, as it is inherently designed for sequential decision making, and can be applied to optimal control problems without the need for a plant system model. To advance education of control engineers and operators in this field, this contribution targets an RL framework that can be applied to educational hardware provided by the Lucas-N\"ulle company. Specifically, the goal of inverted pendulum control is pursued by means of RL, including both, swing-up and stabilization within a single holistic design approach. Herein, the actual learning is enabled by separating corresponding computations from the real-time control computer and outsourcing them to a different hardware. This distributed architecture, however, necessitates communication of the involved components, which is realized via CAN bus. The experimental proof of concept is presented with an applied safeguarding algorithm that prevents the plant from being operated harmfully during the trial-and-error training phase.
\end{abstract}

%\begin{IEEEkeywords}
%\end{IEEEkeywords}

\IEEEpeerreviewmaketitle

\section{Introduction}
\begin{figure*}[htb]
    \centering
    \includegraphics[width=1.0\linewidth]{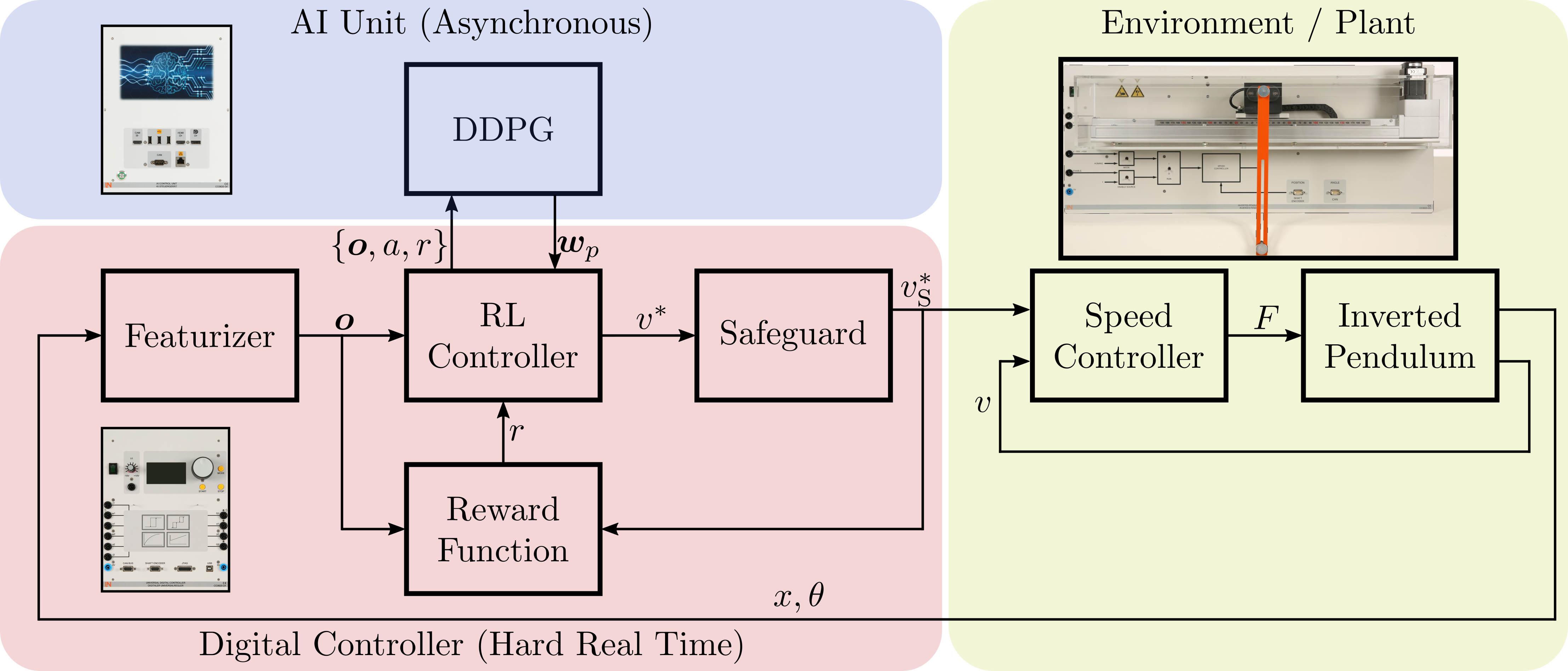}
    \caption{Schematic depiction of the closed-loop control system}
    \label{fig:block_diagram}
\end{figure*}

\IEEEPARstart{T}{his} report highlights the key aspects of implementing a controller for the inverted pendulum problem as a reinforcement learning (RL) application with the use of Lucas-N\"ulle educational hardware. The inverted pendulum, being one of the standard problems of control theory and practice, is usually one of the first practical examples that students get to experiment with. Herein, its intuitiveness, replicability and smooth transition from nonlinear to linear dynamics earned it a special place in automatic control education. While a variety of established control methods such as linearized PID control \cite{Pendulum_PID} or model predictive control \cite{Pendulum_MPC} are available for this nonlinear control plant, this article focuses the implementation of control via RL, which is a subdiscipline of machine learning that corresponds to optimal control. A schematic overview of the employed control concept is depicted in \figref{fig:block_diagram}.

In the following, the discussed key points include the problem setting, the fundamentals of RL-based control and their application to the inverted pendulum. Targeting experimental validation, it is further investigated how to avoid infeasible operation during trial-and-error training (so-called safeguarding) and how the implementation of the overall algorithm on Lucas-N\"ulle hardware is realized. Finally, experimental results from training and application are presented.

\section{Problem Setting}
The inverted pendulum is one of the most studied dynamical systems \cite{Oldest_Pendulum}. Despite being oftentimes viewed a toy example, it actually has a quite practical background, which is underlined by applications like the segway \cite{Segway_Pendulum} or reusable launch vehicles (i.e., safely landing rocket stages) \cite{Rocket_Pendulum}. 

\begin{figure}[htb]
    \centering
    \includegraphics[width=0.8\linewidth]{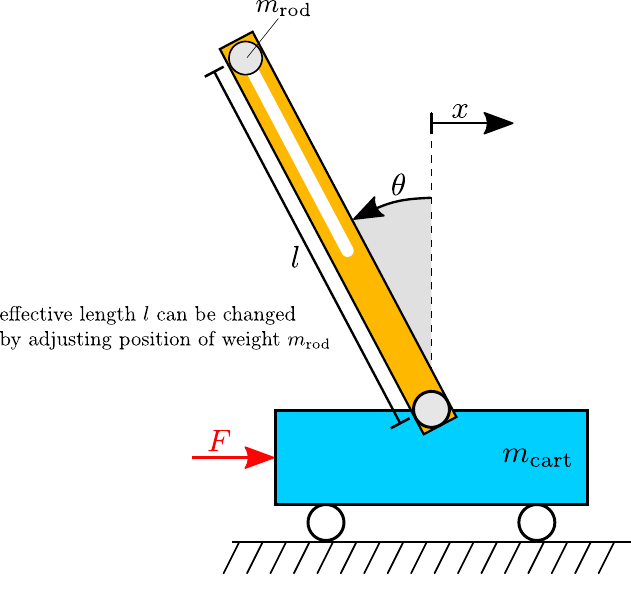}
    \caption{Conceptual depiction of the inverted pendulum}
    \label{fig:free_cut}
\end{figure}

In this investigation, the swing-up and stabilization problem of the inverted pendulum is examined with use of the educational hardware from the Lucas-N\"ulle product portfolio. Herein, the basic scenario of a single pendulum rod on a cart is analyzed, whose schematic is depicted in \figref{fig:free_cut}. Instead of a conventional, model-based control design procedure, the given control problem is tackled by means of RL. The controller design process and resulting controller performance comes with some characteristic traits that are different from conventional control methods:
\begin{itemize}
    \item The controller behavior is adapted in a learning phase. Herein, data from the plant system is collected within direct interaction, allowing consideration of comprehensive dynamics and parasitic effects as long as they are visible within the measurements.
    
    \item Consequently, the controller is not dependent on system modeling. Especially, parameter values do not have to be available, and it is not necessary to identify them.
    
    \item Making use of artificial neural networks (ANNs), a single, nonlinear control law is sufficient to handle both, the swing-up maneuver and the stabilization in the upper equilibrium\footnote{Contrary, linear controllers can only be used close to the equilibrium and conventional optimal controllers usually employ a gain scheduling approach wherein the swing-up is handled differently than the steady state.}.

    \item The learning phase is initialized with an untrained RL controller and, hence, control performance is usually insufficient at the beginning of the training and improves procedurally.
\end{itemize}
This technical report targets to cover theory and implementation of the proposed approach, and will also discuss some experimental results.

Since the modeling and parameterization of the plant system is not of importance for the controller design, it is only briefly revisited for the readers convenience. The inverted pendulum on a cart can be described by a system of differential equations:
\begin{align}
\begin{split}
    F(t)&=
    (m_\text{cart}+m_\text{rod})\frac{\text{d}^2}{\text{d}t^2}x(t)
    - m_\text{rod} l \,\text{cos}(\theta(t)) \frac{\text{d}^2}{\text{d}t^2} \theta(t)
    \\
    &\qquad \qquad
    + m_\text{rod} l\,\text{sin}(\theta(t))\left(\frac{\text{d}}{\text{d}t}\theta(t)\right)^2,
    \\
    0&=   
    l\frac{\text{d}^2}{\text{d}t^2} \theta(t)
    - \cos(\theta (t)) \frac{\text{d}^2}{\text{d}t^2}x(t) 
    - g \sin(\theta(t)),
    \label{eq:model1}
\end{split}
\end{align}
with time $t$, linear position $x$ and gravitational acceleration $g$. The angular displacement $\theta$ takes herein a special role, as the main control goal is the minimization of the absolute displacement $|\theta|$. The mass of the cart and the applied force are denoted by $m_\text{cart}$ and $F$, respectively. The pendulum rod carries a mass $m_\text{rod}$ at its end while the rod itself is assumed weightless. The effective length $l$ of the rod can be adjusted to allow experiments with different moments of inertia. Please note that frictional force is omitted within this model. 

The utilized inverted pendulum experimental rig from Lucas-N\"ulle comes with a readily-tuned speed controller. This can be seen in \figref{fig:block_diagram}, and allows simplification of the mechanical system model when assuming the speed control loop to be significantly faster than the angular movement of the pendulum. Hence, \eqref{eq:model1} collapses to
\begin{align}
\begin{split}
    \frac{\text{d}}{\text{d}t}x(t)&=v(t)\approx v^*(t),
\\
    \frac{\text{d}^2}{\text{d}t^2} \theta(t)&=   
    \frac{1}{l} \cos(\theta (t)) \frac{\text{d}}{\text{d}t}v^*(t) 
    + \frac{g}{l} \sin(\theta(t)),
    \label{eq:model2}
\end{split}
\end{align}
with $v^*$ denoting the speed reference, which herein takes the role of the manipulated variable by means of the RL controller. The momentary system state is specified by the state vector $\bm{s}$
\begin{align}
    \bm{s}(t)=\begin{bmatrix}
        x(t) & v(t) & \theta(t) & \omega(t)
    \end{bmatrix},
\end{align}
whereas $\omega(t)=\frac{\text{d}}{\text{d}t}\theta(t)$ denotes the angular velocity. While the positional states $x$ and $\theta$ are directly measurable from the experimental rig, the velocity states $v$ and $\omega$ are not directly measured and need to be estimated on the basis of the available positional quantities. In the following, availability of corresponding estimate values $\hat{v}$ and $\hat{\omega}$ is assumed. The employed estimation method is presented in Appendix \ref{sec:state_estimation} and subjects to the limiting assumption of no plant parameter knowledge.

\section{Reinforcement Learning Control}
While the core of each RL control approach lies within the proper encoding of the control goal within the reward function, a brief overview over the applied algorithm, the deep deterministic policy gradient (DDPG) is to be presented first \cite{DDPG_Original}.
\subsection{Deep Deterministic Policy Gradient (DDPG)}
The goal of each RL application is the maximization of the return $g$, which is defined as the accumulated future reward:
\begin{align}
    g_k = r_{k+1} + \gamma r_{k+2} + \gamma^2 r_{k+3} +
     \gamma^3 r_{k+4} + \dots.
\end{align}
Herein, the reward function $r$ defines a rating of the momentary performance in terms of the control goal for each time step independently, and the discount factor $\gamma\in[0,1[$ ensures numerical existence of $g$ and allows to prioritize between the near and the far future. 

Further, the action-value function $q$ establishes a relation between the system observation $\bm{o}$, the applied action $a$ and the resulting return $g$:
\begin{align}
    q(\bm{o}_k, a_k)&=g_k=r_{k+1} + \gamma r_{k+2} + \gamma^2 r_{k+3}+\dots.
    \label{eq:action_value}
\end{align}
This holds if and only if the observation-action tuple $(\bm{o}_k,a_k)$ satisfies the Markov property:
\begin{align}
    \bm{o}_{k+1}&=\bm{f}(\bm{o}_k, a_k)
    =\bm{f}(\bm{o}_k, \bm{o}_{k-1}, \dots, a_k, a_{k-1}, \dots),
\end{align}
which means that $(\bm{o}_k,a_k)$ must contain all necessary information that determine the successor observation $\bm{o}_{k+1}$. Please note that the corresponding dynamic function $\bm{f}(\cdot)$ does not need to be known.

Hence, if $q$ would be available, the control action $a^*$ that maximizes $g$ could be identified via
\begin{align}
    a^*_k&=\underset{a}{\text{arg}\,\text{max}}\,q(\bm{o}_k, a),
    \label{eq:argmax}
\end{align}
for each individual observation $\bm{o}$, and the optimal control problem would be solved. From this, the two main tasks for the DDPG can be motivated:
\begin{enumerate}
    \item The action-value function $q$ must be learned to gain access to the relation between optimal return $\text{max}(g)$ and applied action $a^*$, and
    \item the maximization step in \eqref{eq:argmax} must be learned to be performed quickly, because real-time capable applications do not generally permit nonlinear optimization for $a^*$ at runtime.
\end{enumerate}
Note that these tasks basically apply to all RL algorithms that operate on a continuous state and action space. In the following, only the working principles of the well-established DDPG algorithm are discussed. Other eligible algorithms for this task could be, e.g., TRPO, PPO, TD3, SAC \cite{TRPO, PPO, TD3, SAC}, which are not discussed in the scope of this contribution.

\subsubsection{Estimating $q$}
For most applications - and especially for nonlinear systems such as the inverted pendulum - it must be assumed that $q$ is nonlinear. To allow approximation of (strongly) nonlinear functions, deep artificial neural networks (ANNs) have proven themselves as an appropriate tool. Therefore, a feedforward ANN $\hat{q}_{\bm{w}_q}$ is employed as a function approximator for $q$, with corresponding network weights $\bm{w}_q$. Herein, the newly arising task is to adapt these network weights such that the action-value estimate is as accurate as possible, i.e., $\hat{q}_{\bm{w}_q}\approx q=g$. 

Weight adaption is the core of the training routine, and is commonly performed by gradient descent on a cost function. The cost function $J_q$ that is used to compute the update to $\bm{w}_q$ must therefore encode the goal of accurate approximation $\hat{q}_{\bm{w}_q}\approx q$, which can not be done directly because the 'true' action-value function $q$ is unknown. Therefore, $J_q$ must be defined to encode this goal indirectly. Reviewing the definition of $q$ in \eqref{eq:action_value} yields that
\begin{align}
\begin{split}
    q(\bm{o}_k, a_k)&=r_{k+1}+\gamma r_{k+2} + \gamma^2 r_{k+3} +  \gamma^3 r_{k+4}\dots\\
    &=
    r_{k+1}+\gamma \left(
    r_{k+2} + \gamma r_{k+3} + \gamma^2 r_{k+4} \dots
    \right)\\
    &=
    r_{k+1}+\gamma q(\bm{o}_{k+1}, a_{k+1}),
    \label{eq:bellman}
\end{split}
\end{align}
which is popularly known as the Bellman equation. Naturally, also the employed approximation $\hat{q}_{\bm{w}_q}$ should (approximately) satisfy this equation after the training phase:
\begin{align}
    \hat{q}_{\bm{w}_q}(\bm{o}_k, a_k) \approx r_{k+1} +\gamma \hat{q}_{\bm{w}_q}(\bm{o}_{k+1}, a_{k+1}).
\end{align}
Since both sides of this equation only feature available quantities, these can be used to define the cost function to \mbox{determine $\bm{w}_q$}:
\begin{align}
    J_q = \left(
        \hat{q}_{\bm{w}_q}(\bm{o}_k, a_k)
        \right.
        -
        \underbrace{
        \left(
        r_{k+1} +\gamma \hat{q}_{\tilde{\bm{w}}_q}(\bm{o}_{k+1}, a^*_{k+1})
        \right)
        }_{\text{estimation target}}
    \left.
    \vphantom{\hat{q}_{\bm{w}_q}(\bm{o}_k, a_k)}    
    \right)^2.
    \label{eq:cost_q}
\end{align}
From this, the gradient-descent update rule for network weights evaluates to
\begin{align}
    \bm{w}_q \leftarrow \bm{w}_q - \beta_q \nabla_{\bm{w}_q}J_q.
\end{align}
Update rules that employ cost functions of the form \eqref{eq:cost_q} are denoted as bootstrapping procedures for their characteristic of updating an estimator using its very own estimation.
Please note that the estimation target in \eqref{eq:cost_q} is computed making use of target weights $\tilde{\bm{w}}_q$ and assuming the optimal action in the next step $a^*_{k+1}$. Target weights are a delayed version of the original weights, which are usually determined by applying a low-pass filter constant $\kappa\in]0,1[$ such that
\begin{align}
    \tilde{\bm{w}}_q \leftarrow (1-\kappa) \tilde{\bm{w}}_q + \kappa \bm{w}_q,
\end{align}
which has been found to stabilize the training process \cite{VideoGameQ}. The optimal action $a^*$ is yet unavailable and must be approximated, which is addressed in the following.

\subsubsection{Estimating $a^*$}
As denoted by \eqref{eq:argmax}, the optimal action $a^*$ for a given observation $\bm{o}$ could be found by solving a nonlinear maximization problem on $\hat{q}_{\bm{w}_q}$. While this is theoretically possible in a numerical fashion, it is usually infeasible for real-time capable control applications due to its computational burden. Again, a feedforward ANN $\hat{p}_{\bm{w}_p}$ is employed to encode the control policy:
\begin{align}
    \hat{a}_k = \hat{p}_{\bm{w}_p}(\bm{o}_k) \approx a^*_k = \underset{a}{\text{arg}\,\text{max}}\,q(\bm{o}_k, a).
    \label{eq:policy}
\end{align}
Note that the policy $\hat{p}_{\bm{w}_p}$ can be viewed as a nonlinear state feedback with corresponding network weights $\bm{w}_p$. Naturally, also these weights have to be adapted during the training process and the cost function can be directly inferred from \eqref{eq:argmax}. Assuming that the action-value estimate $\hat{q}_{\bm{w}_q}$ is sufficiently accurate, the cost function for the policy $J_p$ can be defined as
\begin{align}
    J_p = -\hat{q}_{\bm{w}_q}(\bm{o}_k, \hat{p}_{\bm{w}_p}(\bm{o}_k))
\end{align}
leading to
\begin{align}
\begin{split}
    \bm{w}_p &\leftarrow
    \bm{w}_p - \beta_p \nabla_{\bm{w}_p} J_p
    \\
    \Leftrightarrow \quad \bm{w}_p &\leftarrow
    \bm{w}_p + \beta_p \nabla_{\bm{w}_p} \hat{q}_{\bm{w}_q}(\bm{o}_k, \hat{p}_{\bm{w}_p}(\bm{o}_k)).
    \label{eq:policy_update}
\end{split}
\end{align}
As the last line of \eqref{eq:policy_update} reveals, minimization of the cost $J_p$ actually corresponds to maximizing the action-value estimate $\hat{q}_{\bm{w}_q}$ and, hence, the expected return $g$. At this point, the algorithm is basically complete and both approximators $\hat{q}_{\bm{w}_q}$ and $\hat{p}_{\bm{w}_p}$ can be trained over time with the use of state transition experiences
\begin{align}
\mathcal{E}=\{\bm{o}_k, a_k, r_{k+1}, \bm{o}_{k+1}\},    
\end{align}
that have to be collected as measurements in interaction with the actual plant system.
For their separation of tasks into state evaluation and action selection, the approximators $\hat{q}_{\bm{w}_q}$ and $\hat{p}_{\bm{w}_p}$ are commonly also denoted as critic and actor, respectively. While the DDPG algorithm is only one of many possible RL algorithms for the task at hand, the alternative algorithms usually feature a very similar, so-called actor-critic structure.

The ANNs that are utilized within this investigation follow a pure feedforward structure (i.e., the ANN is memoryless), which is prominently labeled multilayer perceptron architecture. Accordingly, all network layers are so-called dense layers, characterized by
\begin{align}
    \bm{y}_{i+1}=f_\text{act}(\bm{P}\bm{y}_{i}+\bm{b}),
\end{align}
with layer input $\bm{y}_i$, layer output $\bm{y}_{i+1}$, and layer parameters $\bm{P}$ and $\bm{b}$ (which are condensed within the parameter vector $\bm{w}_p$ for the actor and within $\bm{w}_q$ for the critic). As activation function $f_\text{act}$, this investigation employs the LeakyReLU function for hidden layers, and linear activation for output layers\footnote{Note that $f_\text{act}$ is a scalar function. The notation $f_\text{act}(\bm{y})$ corresponds to element-wise application of $f_\text{act}$ with concern to the components of the vector $\bm{y}$}:
\begin{align}
    \bm{y}_{i+1}&=
    \left\{
    \begin{matrix*}[l]
        \bm{P}\bm{y}_{i}+\bm{b}
        &    
        \text{in the last layer},
        \\
        \text{LeakyReLU}(\bm{P}\bm{y}_{i}+\bm{b})
        &
        \text{otherwise},
    \end{matrix*}
    \right.
\end{align}
with
\begin{align}
    \text{LeakyReLU}(y) = \text{max}(y, \alpha y),
\end{align}
with the leakage parameter being configured to $\alpha=0.3$.

\subsection{Exploration Noise}
To discover and memorize a sensible control strategy, sufficient exploration of the state and action space is necessary. While the policy will develop over the course of the training phase, the employed learning rate $\beta_p$ is usually too low to introduce significant novelty into the collected experiences on its own. Therefore, the policy $\hat{p}_{\bm{w}_p}$ is superimposed with a noise signal $n$ during the training phase:
\begin{align}
    \hat{a}&=
    \left\{
    \begin{matrix*}[l]
        \hat{p}_{\bm{w}_p}(\bm{o}_k)
        &
        \text{during application},
        \\
        \hat{p}_{\bm{w}_p}(\bm{o}_k) + n_k
        &
        \text{during training}.
    \end{matrix*}
    \right.
\end{align}

To a limited extent, this so-called exploration noise adds randomness to the selected actions and, hence, to the state transitions that are seen during training, enabling the DDPG to consider control commands that are unlike the current policy.

While basically any random process could be applied, it has been found that Ornstein-Uhlenbeck (OU) noise is a solid choice for inert systems \cite{DDPG_Original}. It has the form
\begin{align}
    n_k &=
    (1-\mu T_\text{s})n_{k-1}  + \sigma \sqrt{T_\text{s}} \mathcal{N}_{(0,1)},
\end{align}
with the mean reversion rate $\mu$ and diffusion factor $\sigma$. The symbol $\mathcal{N}_{(0,1)}$ denotes a random sample from a normal distribution with mean $0$ and variance $1$. As can be seen, OU noise is a stateful process, making it better suited to excite systems with significant low-pass behavior. However, the newly introduced parameters $\mu$ and $\sigma$ also result in more tuning effort. 

A structured overview of the DDPG algorithm is provided in Alg. \ref{alg:pseudocode}. Herein $\mathcal{B}$ denotes a minibatch of several experiences which are randomly drawn from a memory buffer $\mathcal{M}$ that holds a multitude of experience tuples. In this investigation, $\mathcal{M}$ is designed as a circular buffer, meaning that oldest experiences are first to be overwritten.

\begin{algorithm}
\caption{Deep Deterministic Policy Gradient (DDPG)}
\label{alg:pseudocode}
\begin{algorithmic}
\State Randomly initialize weights of $\hat{q}_{\bm{w}_q}$ 
and $\hat{p}_{\bm{w}_p}$
\State Initialize target weights accordingly: 
\State \mbox{$\tilde{\bm{w}}_q \leftarrow \bm{w}_q$} and \mbox{$\tilde{\bm{w}}_p \leftarrow \bm{w}_p$}
\Repeat
\State Measure $\bm{s}_k$
\State Compute $\bm{o}_k$
\State Compute $\hat{a}_k = \hat{p}_{\bm{w}_p}(\bm{o}_k)$
\If{Training}:
\State Superimpose noise $\hat{a}_k \leftarrow \hat{a}_k + n_k$
\EndIf
\State Execute $\hat{a}_k$ and measure $r_k$ and $\bm{s}_{k+1}$, compute $\bm{o}_{k+1}$
\State Save $\mathcal{E}$ to memory: $\mathcal{M} \leftarrow \{\bm{o}_k, \hat{a}_k, r_{k+1}, \bm{o}_{k+1}\}$
\State Sample random experience batch $\mathcal{B} \subset \mathcal{D}$
\State Compute $J_q$ on $\mathcal{B}$:
\State 
$\quad
J_q = \frac{1}{|\mathcal{B}|}\sum_{\mathcal{E} \in \mathcal{B}}
\left(
        \hat{q}_{\bm{w}_q}(\bm{o}_k, a_k) \right.$
\State $\qquad \qquad
        \left.
        -
        \left(
        r_{k+1} +\gamma \hat{q}_{\tilde{\bm{w}}_q}(\bm{o}_{k+1}, 
        \hat{p}_{\tilde{\bm{w}}_p}(\bm{o}_{k+1}))
        \right)
    \right)^2
    $
\State Compute $J_p$ on $\mathcal{B}$:
\State 
$
\quad J_p = -\frac{1}{|\mathcal{B}|}\sum_{\mathcal{E} \in \mathcal{B}}
    \hat{q}_{\bm{w}_q}(\bm{o}_k,\hat{p}_{\bm{w}_p}(\bm{o}_k))
$
\State Compute and apply the gradient updates:
\State $\quad \bm{w}_q \leftarrow \bm{w}_q - \beta_q \nabla_{\bm{w}_q}J_q$
\State $\quad  \bm{w}_p \leftarrow
    \bm{w}_p - \beta_p \nabla_{\bm{w}_p} J_p$
\State Update weights of target networks:
\State \qquad $\tilde{\bm{w}}_q \leftarrow (1-\kappa) \tilde{\bm{w}}_q + \kappa \bm{w}_q$
\State \qquad $\tilde{\bm{w}}_p \leftarrow (1-\kappa) \tilde{\bm{w}}_p + \kappa \bm{w}_p$
\State $k \leftarrow k+1$
\Until{convergence condition is met}
\end{algorithmic}
\end{algorithm}

\subsection{Observation Design / Feature Engineering}
Coming back to the inverted pendulum, it has already been discussed that the unmeasurable state variables $v$ and $\omega$ can be estimated without any further expert knowledge (cf. Appendix \ref{sec:state_estimation}). Instead of feeding only the measurable states $x, \theta$ into the DDPG algorithm, the Markov property can only be fulfilled by adding these state estimations to the observation vector $\bm{o}$. Further, all available information should be normalized to the range of \mbox{$[-1, 1]$}, which improves the training speed. For the problem at hand. the newly crafted observation vector $\bm{o}$ takes the form
\begin{align}
\begin{split}
    \bm{o}_k = & \left[ 
        \frac{x_k}{x_{\text{max}}} \quad 
        \frac{\hat{v}_k}{v_{\text{max}}} \quad 
        \cos(\theta_k) \quad 
        \sin(\theta_k) 
        \right. 
        \\
        & \quad \quad
        \left.
        \frac{\hat{\omega}_k}{\omega_{\text{max}}} \quad 
        \frac{v^*_{k-1}}{v_{\text{max}}} \quad 
        \frac{x_{\text{ref}, k}}{x_{\text{max}}} \quad
        \frac{x_{\text{ref}, k} - x_k}{2x_{\text{max}}}
        \right].
        \label{eq:obs_design}
\end{split}
\end{align}
Note that the normalization of $\theta$ is herein performed by application of $\cos(\cdot)$ and $\sin(\cdot)$. This choice avoids step-like changes of the angle information\footnote{Usual angular sensors are limited to the range of $[-\pi, \pi]$ or $[0, 2\pi]$.}, which otherwise would be prone to induce step-like changes of the action $\hat{a}$.

While the main goal is the stabilization in the upper equilibrium and, hence, $\theta\approx 0$, the linear positioning $x$ of the rod is yet arbitrary. Therefore, as a secondary goal, the controller should target stabilizing the pendulum in a specific position $x_\text{ref}$, which is to be understood as the reference position for the cart. Naturally, this quantity is specified externally and needs to be added to $\bm{o}$.

Similarly to \eqref{eq:obs_design}, also the network output $\hat{a}$ should be limited to the range of $[-1, 1]$. For the actor, this can be ensured by employing the denormalization
\begin{align}
    v^*_k=\hat{a} v_\text{max}.
\end{align}
For the critic, the range of the network output $\hat{q}_{\bm{w}_q}$ is dependent on the reward design, which is discussed in the following.

\subsection{Reward Design}
\label{sec:reward_design}
The reward design must encode the control goal by quantifying the control performance for the momentary step. Herein, it may employ arbitrarily nonlinear functions, such as case distinctions, which is generally different from conventional optimal control methods \cite{PowerMPC}. Making use of case distinctions, it is possible to define priorities of control goals. Each priority level is assigned to a specified region of the state-action space, and each region's reward targets a distinct controller behavior.

The three regions are distinguished by the identifiers $\mathbb{C}$, $\mathbb{B}$ and $\mathbb{A}$, with $\mathbb{C}$ corresponding to the lowest and $\mathbb{A}$ corresponding to the highest reward. In the following, each of these regions is discussed in detail.

Region $\mathbb{C}$: the RL controller is trying to command a speed $v^*$ that is outside the speed range the pendulum cart is capable of. While harmless in the context of the physical plant, it should be avoided to perpetually operate within the input limitation because few informative experiences can be collected here. Particularly, depending on the severity of input saturation (i.e., how far the control commands lie outside their limitation), the employed exploration noise may be unlikely to compensate for the infeasibly high controller gain. This would nullify exploration, motivating to force the control commands back into the feasible range by means of the reward. Accordingly, the reward is defined to increase as the control command $v^*$ is moved closer to its allowed range $[-v_\text{max}, v_\text{max}]$.
\\
\textbf{If} $v_\text{max} < |v_k^*|$:
\begin{align}
\begin{split}
r_{k+1}&=
    (1-\gamma)
    \left(
        \frac{|v_k^*|-v_\text{max}}{v_\text{max}}-1
        \right),
    \\
    r_{k+1} &\in [-\infty,-(1-\gamma)].
\end{split}
\end{align}

Region $\mathbb{B}$: the controller is operating in the feasible speed range and the actual control task can be tackled. Herein, the pendulum angle $\theta$ is to be moved towards zero. Being defined on the interval $\theta\in[-\pi, \pi]$, the cosine function is the intuitive candidate. 
\\
\textbf{If} $|v_k^*| < v_\text{max}$ \textbf{and} $\theta_\text{thresh} < |\theta_k|$:
\begin{align}
\begin{split}
    r_{k+1}&=
    \frac{1}{4}
    (1-\gamma)
    \left(
        \cos(\theta_k) - 3
    \right),
    \\
    r_{k+1} &\in \left[-(1-\gamma), -\frac{1-\gamma}{2}\right].
    \label{eq:reward_B}
\end{split}
\end{align}

Region $\mathbb{A}$: although maximizing \eqref{eq:reward_B} is a theoretically complete description of the angle control task, the training can be simplified by adding a third priority level. Because a minimum absolute angular speed $|\omega|$ is necessary to move the pendulum into its upper position, the RL agent must firstly learn to increase $|\omega|$ sufficiently to reach $\theta \approx 0$. Without further measures, however, the agent is prone to maintain a high $|\omega|$ to reliably collect a high reward when periodically visiting the upper position. As this conflicts with the control goal of operating constantly in the upper equilibrium, a threshold angle $\theta_\text{thresh}$ is defined, beyond which the reward is increased with decreasing $|\omega|$. Moreover, the absolute positioning error $|x_\text{ref}-x|$ is incorporated only in this region to prioritize angular stabilization over linear positioning, i.e., the positioning task is to be tackled only if stabilization ($\theta\approx 0$) has been successful before.
\\
\textbf{If} $|v_k^*| < v_\text{max}$ \textbf{and} $|\theta_k| < \theta_\text{thresh}$
\begin{align}
\begin{split}
    r_{k+1}&=
    \frac{1}{4}
    \left(
    3
    \left(
        1-\frac{|\hat{\omega}|}{\omega_\text{safe+}}
    \right)
    \right.
    \\
    & \quad \quad
    \left.
    +
    3
    \left(
        1-\frac{|x_\text{ref}-x|}{2 x_\text{max}}
    \right)
    -2
    \right)
    (1-\gamma),
    \\
    r_{k+1} &\in \left[-\frac{1-\gamma}{2}, (1-\gamma)\right].
\end{split}
\end{align}

\begin{figure}[htb]
    \centering
\includegraphics[width=1.0\linewidth]{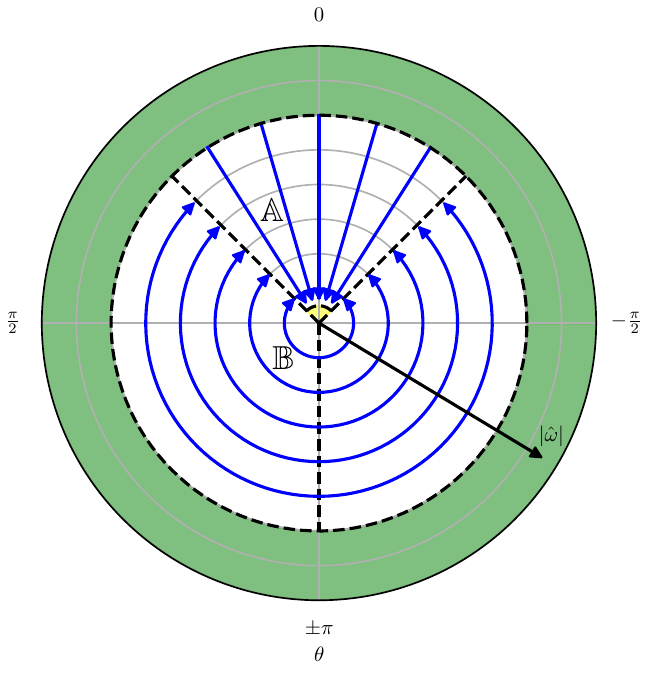}
    \caption{Depiction of the reward gradients with respect to angle $\theta$ and angular velocity $\hat{\omega}$, the green shaded area corresponds to safeguard activation, the yellow shaded area corresponds to optimal operation}
    \label{fig:reward_gradient}
\end{figure}

A visual representation of the reward can be gained from its gradient with respect to the state. For the given problem, such a depiction is provided in \figref{fig:reward_gradient} for the angle-related states $\theta$ and $\hat{\omega}$. Herein, the reward regions $\mathbb{B}$ and $\mathbb{A}$ are of interest.

\section{Safeguarding}
In real-world experiments, RL training may come with a safety risk due to the randomly initialized control policy and the superimposed exploration noise. For that, safeguarding measures can be employed to protect the plant system from damage and avoid emergency shutdowns (and therefore downtime). In simple control environments, safeguarding can even be designed to exclude state-space regions with unintended characteristics from the training, allowing the RL agent to spend more time in the state-space regions of interest.

As the Lucas-N\"ulle inverted pendulum is an educational device, it is constructed with the specific requirement to be robust with concern to infeasible control commands. As an inherent protective measure, the plant will shut down when a positional limit is violated $|x|>x_\text{max}$. This would require a safe reset by the user, which is undesired for the targeted training phase, as it should run fully autonomously. Therefore, as a first safeguarding measure, the pendulum cart is steered to the center position $x=0$ whenever it gets too close to either positional limit:
\newpage
\begin{algorithmic}
    \If{$|x+3T_\text{s}v^*|\geq x_\text{max}$}
        \Repeat
            \State $v^*_\text{S} \leftarrow -\text{sign}(x)v_\text{max}$
        \Until{$x \approx 0$}
    \EndIf
\end{algorithmic}
Further, as discussed in \mbox{Sec. \ref{sec:reward_design}}, a too great increase of the absolute angular velocity $|\omega|$ might be problematic for the training. To force $\omega$ back into feasible regions, all control commands are nullified $v^*=0$, which means that no further energy is fed into the plant:
\begin{algorithmic}
    \If{$|\hat{\omega}| \geq \omega_\text{safe+}$}
        \Repeat
            \State $v^*_\text{S} \leftarrow 0$
        \Until{$|\hat{\omega}| \leq \omega_\text{safe-}$}
    \EndIf
\end{algorithmic}
Due to mechanical friction in the rod's bearing, the angular velocity will decrease over time and the usual RL control can be continued as soon as $|\omega|$ is low enough again. Please note that the bearing friction is a parasitic effect that has not been incorporated into the plant model \eqref{eq:model1}.

\section{Implementation}
The whole hardware implementation, spanning the pendulum as plant system, a digital signal processor (DSP) and an edge-computing device (ECD), comes from the Lucas-N\"ulle product portfolio. Their corresponding brand names and unique identifiers are listed in \tabref{tab:devicelist}, and the physical parameters of the plant are stated in \tabref{tab:symbollist}. Please note that the specification of the pendulum parameters ($m_\text{rod},m_\text{cart},l$) is herein only delivered for the readers convenience. They where not used in the configuration or training of the RL controller.

\begin{table}[htb]
\centering
\begin{tabular}{l l l}
\toprule
\textbf{Brand Name} & \textbf{Description} & \textbf{Identifier} \\
\hline
Inverted Pendulum & Plant System & CO3620-2G\\
Universal Digital Controller & DSP & CO3620-2A \\
AI Control Unit & ECD & CO3620-3A\\
\bottomrule
\end{tabular}
\caption{Components of the Lucas-N\"ulle test bench system}
\label{tab:devicelist}
\end{table}

The overall computational burden can be distributed between the DSP and the ECD: while the actual control routine in form of the actor network $\hat{p}_{\bm{w}_p}$ must be deployed with real-time capability on the DSP, the learning procedure for its weights can be executed asynchronously on the ECD. For the latter, no real-time constraint applies. 

\begin{figure}[htb]
    \centering
    \includegraphics[width=1.0\linewidth]{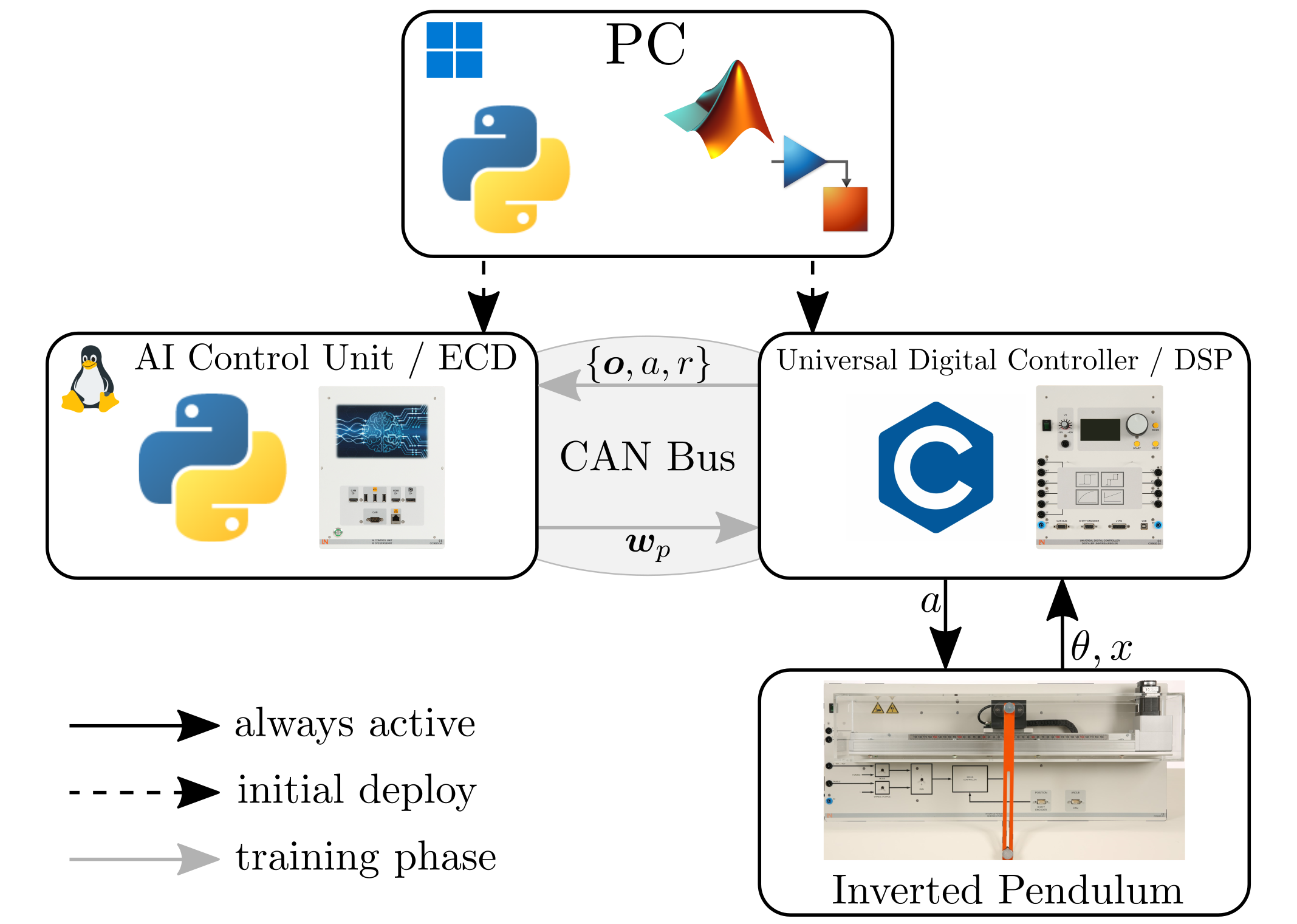}
    \caption{Schematic of the employed online RL setup}
    \label{fig:CAN_implementation}
\end{figure}

To realize this separation of tasks, the information in question has to be communicated between the two devices. Specifically, the DSP process has to acquire and send the information concerning state transition experiences $\mathcal{E}$, and the ECD, which is concerned with the actual RL training, updates the actor weights $\bm{w}_p$ and sends them procedurally to the DSP (cf. \figref{fig:block_diagram}). Notably, the critic network $\hat{q}_{\bm{w}_q}$ is not needed for the real-time control routine and, hence, it can be kept entirely inside the ECD. The two-way communication is set up via CAN bus and is operated on a bandwidth of $1 \,\text{Mbit}/\text{s}$. An overview of the controller and communication setup is provided by \figref{fig:CAN_implementation}.

The learning routine on the ECD is strongly inspired by \cite{EdgePaper}. It is implemented within Python and utilizes the libraries kerasRL \cite{KerasRL} and Tensorflow \cite{TensorflowPaper} for machine learning. The DSP is programmed via a compiled deploy from MATLAB - Simulink \cite{MATLAB}. The real-time condition that applies to the DSP is the overall bottleneck of the given setup, as it limits the allowed complexity of $\hat{p}_{\bm{w}_p}$. The underlying computation has to be finished after a maximum of \mbox{$T_\text{s} = 1/f_\text{s} = 20 \,\text{ms}$}. The computational power of the ECD is not subjected to such conditions. Yet, it affects the duration of the training process because the parameter updates come with high calculation effort and the convergence of the control performance is accelerated by increasing the update rate. For the given setup, a training phase of $T_\text{t}=30\,\text{min}$ is configured.

Over the course of the training time, the learning rates of actor and critic are decreased in a linear fashion from $\beta^\text{init}$ to $\beta^\text{fin}$. Likewise, the exploration variance $\sigma$ is decreased over time to allow a smooth transition from exploration- to performance-oriented collection of data. 

\begin{table}[htb]
\centering
\begin{tabular}{l l l r l}
\toprule
\multicolumn{2}{l}{\textbf{Symbol}} & \textbf{Description} & \textbf{Value} \\
\hline
\multirow{4}{*}{\rotatebox[origin=c]{90}{Plant}} 
& $m_\text{rod}$ & Mass of the Pendulum Weight & $0.3$ & kg\\
& $m_\text{cart}$ & Mass of the Pendulum Cart & $0.865$ & kg\\
& $l$ & Effective Pendulum Length & $[0.135, 0.29]$ & m\\
& $f_\text{s}$ & Sampling Frequency & $50$ & Hz\\
\cline{1-1}
\multirow{8}{*}{\rotatebox[origin=c]{90}{Control Framework}}
& $x_\text{max}$ & Maximum Linear Position & $0.2$ & m\\
& $x_\text{safe}$ & Safeguarded Linear Position & $0.17$ & m\\
& $v_\text{max}$ & Maximum Linear Velocity & $0.5$ & $\frac{\text{m}}{\text{s}}$\\
& $\theta_\text{thresh}$ & Angle-Minimization Threshold & $\frac{\pi}{4}$ & \\
& $\omega_\text{safe+}$ & Safeguarded Upper Limit for $\omega$ & $6 \pi$ & $\frac{1}{\text{s}}$\\
& $\omega_\text{safe-}$ & Safeguarded Lower Limit for $\omega$ & $\frac{1}{10} \pi$  & $\frac{1}{\text{s}}$\\
& $ f_0 $ & PLL Bandwidth & $7$ & Hz\\
& $ d$ & PLL Damping Factor & $1$ &\\
\cline{1-1}
\multirow{16}{*}{\rotatebox[origin=c]{90}{DDPG}}
& $T_\text{t}$ & Duration of Training Phase & $30$ & $\text{min}$\\
& $\gamma$ & Discount Factor & 0.95 & \\
& $\alpha$ & Leakage Parameter & $0.3$ & \\
& & Critic Layers & $5$ & \\
& & Critic Neurons per Hidden Layer & $200$ & \\
& $\beta_q^\text{init}$ & Initial Critic Learning Rate & $1\cdot 10^{-3}$ & \\
& $\beta_q^\text{fin}$ & Final Critic Learning Rate & $1\cdot 10^{-4}$ & \\
& & Actor Layers & $3$ & \\
& & Actor Neurons per Hidden Layer & $128$ & \\
& $\beta_p^\text{init}$ & Initial Actor Learning Rate & $2.5\cdot 10^{-3}$ & \\
& $\beta_p^\text{fin}$ & Final Actor Learning Rate & $2.5\cdot 10^{-4}$ & \\
& $\kappa$ & Target Netork Update Parameter & $15\cdot 10^{-2}$ &  \\
& $|\mathcal{B}|$ & Batch Size & $64$ &  \\
& $|\mathcal{M}|$ & Memory Size & $6\cdot 10^4$ &  \\
& $\mu$ & OU Reversion Rate & $2$ &  \\
& $\sigma^{\text{init}}$ & Initial OU Diffusion Factor & $0.2$ &  \\
& $\sigma^{\text{fin}}$ & Final OU Diffusion Factor & $0$ &  \\
\bottomrule
\end{tabular}
\caption{Specification of the inverted pendulum control system}
\label{tab:symbollist}
\end{table}

\section{Experimental Results}
The proposed setup is to be analyzed concerning its convergence behavior during training and concerning its closed-loop performance. All results are collected in real-world experiments. The secondary goal of linear positioning is herein simplified by defining $x_\text{ref}= 0$ throughout all experiments.

\subsection{Training Phase}
For investigation of the training phase, it is of interest how reliable the control performance converges after the specified training time of $T_\text{t}=30\,\text{min}$. For that, a set of ten independent control agents have been trained on the inverted pendulum, whereas each of these agents has been initialized randomly. From these ten training runs, five are conducted with the pendulum weight in the outer position ($l\approx0.29\,\text{m}$), and the other five with the weight in the inner position ($l\approx0.135\,\text{m}$). The overall algorithm was not altered to allow a conclusion on the generalizability of the full setup.

\begin{figure*}[htb]
    \centering
 \includegraphics[width=1.0\linewidth]{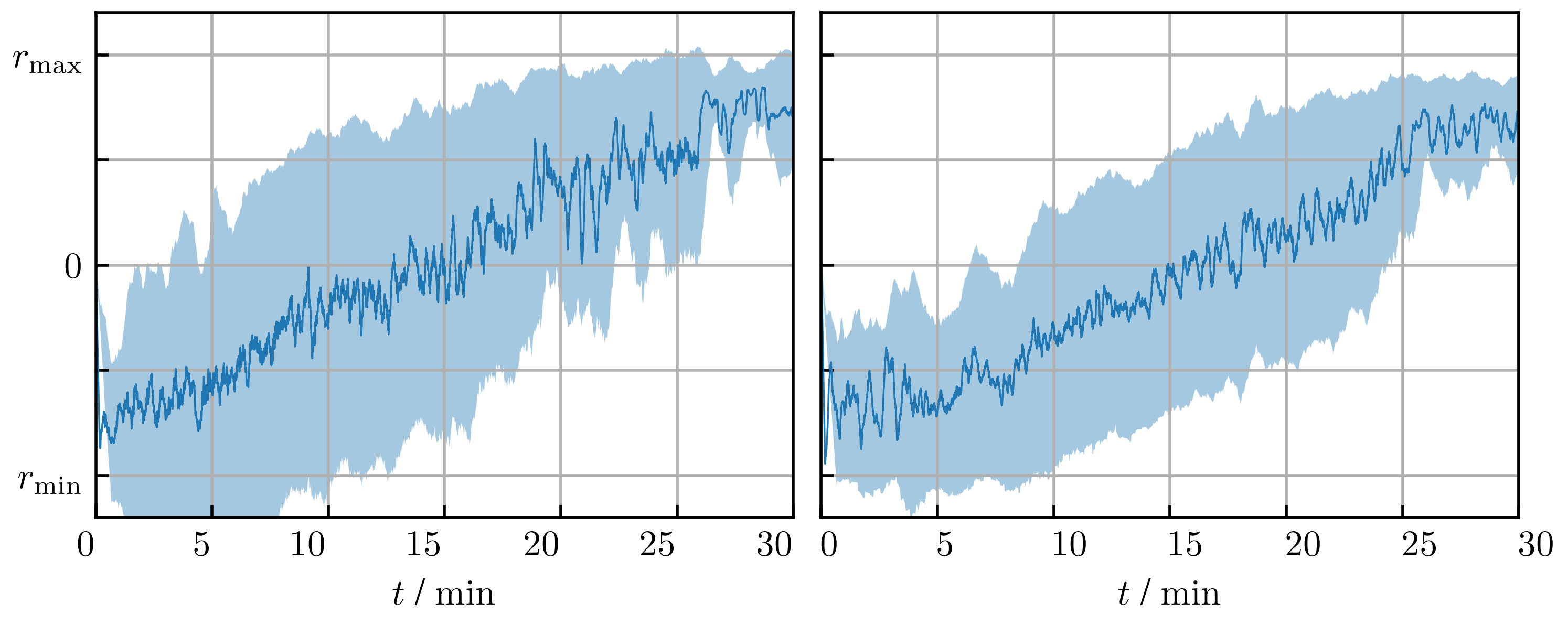}
    \caption{Left: average convergence behavior of five different RL controllers trained with the pendulum weight in the innermost position, Right: average convergence behavior of five different RL controllers trained with the pendulum weight in the outermost position; the blue shaded area denotes one standard deviation}
    \label{fig:convergence}
\end{figure*}

The convergence behavior is depicted in \figref{fig:convergence}. Herein, the learning progress can be seen clearly for both of the two cases. Whereas the standard deviation is quite striking up to $t\approx 25\,\min$, it decreases significantly for the last 5 minutes, denoting a quite reliable convergence. Most importantly, no significant difference in convergence behavior can be observed for the change from lowest to largest effective length $l$, indicating the independence of the RL control approach from specific parameters and their availability.

All ten agents were finally capable of a swing-up maneuver, with only a single controller not being capable of stabilizing the pendulum in the upper equilibrium indefinitely. Moreover, as the average reward curve is not hitting the upper reward boundary $r_\text{max}$ for either case in \figref{fig:convergence}, it can be inferred that the controller did not manage to track the linear position $x_\text{ref}=0$ in all cases. It can be speculated that a further increase of the training time would help to increase the reliability to $100\,\%$ for both of these issues.

\subsection{Application Phase}
\begin{figure}[htb]
    \centering
 \includegraphics[width=1.0\linewidth]{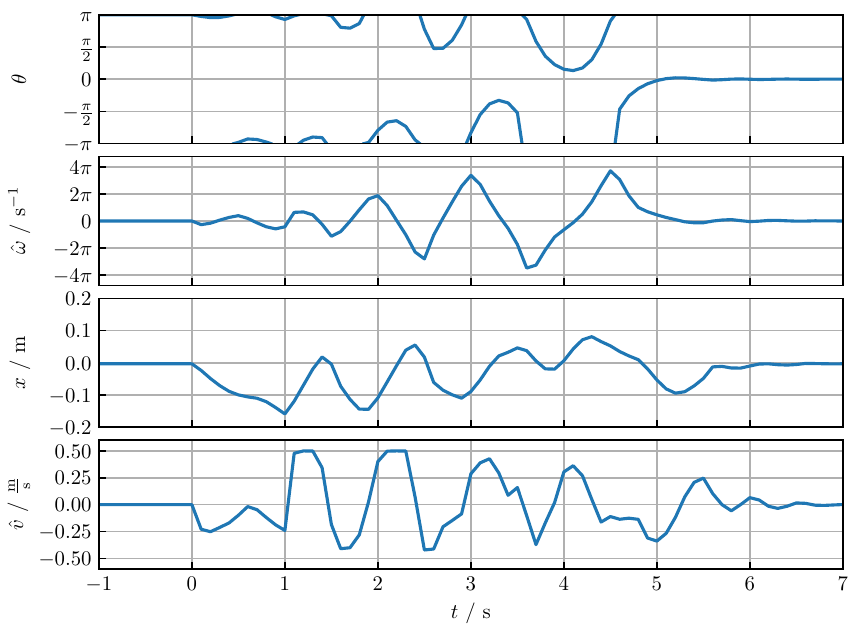}
    \caption{Timeseries plot of an exemplary swing-up maneuver with the trained RL controller. The controller is enabled at $t=0\,\text{s}$ and the weight is in its outermost position.}
    \label{fig:timeseries}
\end{figure}

To conclude on the performance, an exemplary swing-up maneuver is recorded and depicted in \figref{fig:timeseries}. Taking roughly $6.5 \,\text{s}$, the swing-up and stabilization cannot yet be claimed optimal in terms of reaction time. However, the stabilization is successful and the depicted agent seems to track the reference position $x_\text{ref}=0$ with good precision. 

Further, at $t\approx 1\,\text{s}$ it can be seen that the safeguard is activated to prevent the cart from violating the positional limit at $x=-0.2\,\text{m}$. The resulting movement with maximum positive speed is exploited for the swing-up, demonstrating that the agent has learned to utilize the safeguard's behavior for successful operation. Hence, operation of this specific controller without its safeguard cannot be expected to be successful, and a more comprehensive reward design that penalizes safeguard activation would be needed to discourage such behavior.

\section{Conclusion and Outlook}
The proposed RL control pipeline for the swing-up and stabilization of the inverted pendulum has been successfully applied to the educational hardware from Lucas-N\"ulle. After a $30\,\text{min}$ training phase, it can be operated without expert knowledge about the pendulum dynamics and its parameters, rendering the negligence of parasitic effects of frictional force and the rod's moment of inertia uncritical. Only the knowledge about limitations of the system have been utilized to sensibly limit the numerical range of rewards, and to configure the safeguard in order to prevent the plant from undesired behavior that could negatively affect the learning success.

For future extension of the employed control algorithm, it is of interest to increase the convergence speed during the training phase to shorten the overall training time, which could be achieved by comprehensive hyperparamter optimization. This may also allow a better positional tracking behavior, which was not prioritized in this contribution. A reward design that takes the presence of the safeguard into account could allow for a controller that does not rely on the safeguard's intervention. 

\appendix
\subsection{State Estimation}
\label{sec:state_estimation}
In the given setup, the state variables of the cart's linear velocity $v$ and the angular velocity $\omega$ are not directly available through measurement, because the plant is only equipped with sensors for the linear and the angular position $x$ and $\theta$. However, the RL control approach relies on the Markov property to be trainable and, therefore, corresponding state information must be available to render the learning phase successful. Based on the assumption that no parameter information is available for the given system, a classical state observer, e.g., based the Luenberger or Kalman approach \cite{Luenberger, Kalman}, cannot be designed to make state estimations.

Fortunately, the missing state information corresponds to the available measurements of $x$ and $\theta$ with an obvious relation:
\begin{align}
    v(t) &= \frac{\text{d}}{\text{d}t}x(t), & \omega(t) &= \frac{\text{d}}{\text{d}t}\theta(t).
\end{align}
While discrete-time approximation of the derivative (e.g., via the difference quotient) would be an evident way to allow corresponding estimations $\hat{v}$ and $\hat{\omega}$ in the digital control system, this approach is rejected due to its numerical sensitivity. Instead, the linear part of a digital phase-locked loop (PLL) can be employed to compute a smooth and stable estimate of the derivative at runtime \cite{BestPLL}. 

\begin{figure}[htb]
    \centering
    \includegraphics[width=0.8\linewidth]{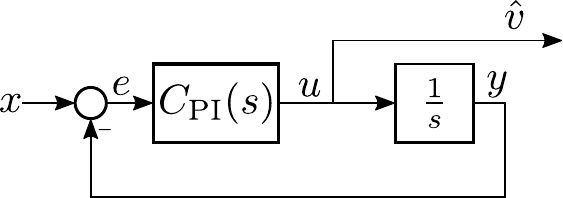}
    \caption{Linear part of a PLL as an estimator for the linear speed $v$ (with corresponding estimate $\hat{v}$), with measured linear position $x$ as input}
    \label{fig:PLL}
\end{figure}

A block diagram for the underlying PLL-based algorithm is provided in \figref{fig:PLL}. As can be seen, it consists of nothing more than an integrator $\frac{1}{s}$ that is 'controlled' by a proportional-integral (PI) element 
\begin{align}
    C_\text{PI}(s)&=K_\text{P} + K_\text{I}\frac{1}{s}.
\end{align} 
This closed-loop structure is purely algorithmic, all its components are virtual and not to be understood as an actual control system with physical states. 

Assuming a steady state, the virtual control error $e$ would be approximately zero:
\begin{align}
\begin{split}
    e(t) &=x(t)-y(t) \approx 0,
    \\
    \Leftrightarrow
    x(t) &\approx y(t) = \int u(t) \text{d}t,
    \\
    \Leftrightarrow
    u(t)&\approx\frac{\text{d}}{\text{d}t}x(t)=v(t).
    \label{eq:locked_PLL}
\end{split}
\end{align}
Hence, the virtual actuator signal $u$ is an estimate $\hat{v}$ for the linear speed $v$ of the pendulum cart that is accessible without any parameter knowledge and can be assumed accurate as long as the PI element $C_\text{PI}$ 'controls' the virtual integrator plant with sufficient precision. To investigate the virtual control error, the input-to-error transfer function is examined in the Laplace domain:
\begin{align}
\begin{split}
    \frac{E(s)}{X(s)}&=\frac{s^2}{s^2+s K_\text{P} + K_\text{I}}
    =\frac{s^2}{s^2+2 d \omega_0 s + \omega_0^2},
    \\
    \text{with}\quad & x(t) \, \laplace \, X(s), \quad e(t)\,\laplace\,E(s).
\end{split}
\end{align}
indicating that $C_\text{PI}$ can be easily tuned by defining the desired bandwidth $\omega_0 = 2 \pi f_0$ and damping factor $d$. This concept and its advantages transfer to the discrete-time domain without loss of generality, meaning that the digital implementation of the PLL structure \figref{fig:PLL} still satisfies \eqref{eq:locked_PLL}.

\begin{figure}[htb]
    \centering
    \includegraphics[width=0.8\linewidth]{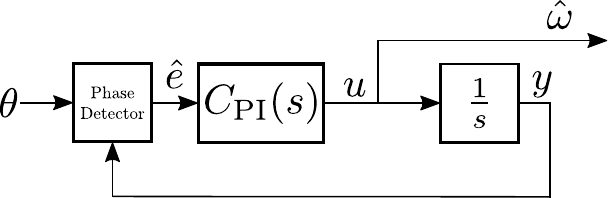}
    \caption{PLL with phase detector as an estimator for the angular speed $\omega$ (with corresponding estimate $\hat{\omega}$), with measured angle $\theta$ as input}
    \label{fig:PD_PLL}
\end{figure}

As angular sensors are oftentimes limited to the range \mbox{$\theta \in [-\pi, \pi]$}, this scheme must be extended to handle discontinuous transition of the measurement signal\footnote{The specific range is not critical for the employed extension. It is therefore also valid for angular sensors with a range of e.g., $\theta \in [0, 2\pi]$.} to be applicable also for estimation of the angular speed $\omega$. A nonlinear phase detector replaces the summation element as depicted in \figref{fig:PD_PLL}.
Herein, the phase detector performs the operation
\begin{align}
    \begin{split}
        \hat{e}(t)&=\sin(\theta(t))\cos(y(t))-\sin(y(t))\cos(\theta(t)),
    \end{split}
\end{align}
which is insensitive to step-wise changes from, e.g., $\theta=-\pi$ to $\theta=\pi$ and will still generate a smooth output $\hat{e}$ in such cases\footnote{Note that $\hat{e}(t)=\sin(\theta(t)-y(t))\approx \theta(t)-y(t)$.}.

In conclusion, the estimation of linear and angular speed $\hat{v}$ and $\hat{\omega}$ can be realized by means of a PLL algorithm, which is computationally cheap and does not require parameter knowledge. The employed selection of PLL parameters $f_0$ and $d$ is specified in \tabref{tab:symbollist}. 

\bibliographystyle{IEEEtran}
\bibliography{main}

%\begin{IEEEbiography}
%[{\includegraphics[width=1in,height=1.25in,clip,keepaspectratio]{Portraits/Maximilian_Schenke.pdf}}]{Maximilian Schenke} received the bachelor's  and master's (hons.) degrees in electrical engineering from Paderborn  University,  Paderborn,  Germany, in 2017 and 2019, respectively. He is currently working toward the doctoral degree in electrical engineering with the Department of Power Electronics and Electrical Drives. His research interests include the application of reinforcement learning methods in the domain of drive control.
%\end{IEEEbiography}
%\begin{IEEEbiography}
%[{\includegraphics[width=1in,height=1.25in,clip,keepaspectratio]{Portraits/Oliver_Wallscheid.pdf}}]{Oliver Wallscheid} (S'13-M'17) received the bachelor's and master's degrees (hons.) in industrial engineering and the doctorate degree (hons.) in electrical engineering from the Paderborn University, Paderborn, Germany, in 2010, 2012, and 2017, respectively. Since 2017, he has been a Senior Research Fellow with the Department of Power Electronics and Electrical Drives at Paderborn University. Currently he is also serving as an acting professor of the Automatic Control Department at Paderborn University. His research focus is on data-driven identification and intelligent control of electrical power systems in decentralized grids, power electronics and drives.
%\end{IEEEbiography}

\vfill

\end{document}